\documentstyle[12pt]{article}
\begin{document}
\vspace{1.cm}
\begin{center}
\    \par
\    \par
\    \par
\     \par
  \bf{ THE  COMPARISON OF CHARACTERISTICS OF $\pi^{-}$ MESONS PRODUCED
IN CENTRAL Mg-Mg INTERACTIONS WITH THE QUARK GLUON STRING MODEL
 PREDICTIONS}
\end{center}

\   \par
\   \par
\   \par
\   \par
\    \par
\par
L.Chkhaidze, T.Djobava,  L.Kharkhelauri, M.Mosidze\par
High Energy Physics Institute, Tbilisi State University,\par
University St 9, 380086 Tbilisi, Republic of Georgia\par
Fax: (99532) 99-06-89;  E-mail: djobava@sun20.hepi.edu.ge or\par
\hspace{6.5cm}                     ida@sun20.hepi.edu.ge \par
\pagebreak
\begin{center}

                     \bf{ ABSTRACT }
\end{center}
\ \par
\par
A detailed study of pion production in central Mg-Mg collisions at a momentum
of 4.3 GeV/c per incident nucleon was carried out with use of the setup GIBS.
The average kinematical characteristics of pions  (multiplicity $n_{-}$
, momentum $P$,
transverse momentum $P_{T}$, emission angle  $\Theta$, rapidity $Y$)
 and corresponding distributions have been obtained.
The experimental results have been compared with the predictions of the Quark
Gluon
String Model (QGSM) and satisfactory  agreement between
 the experimental data and the model has been found.
The QGSM reproduces also the dependence of average $P_{T}$
on  $n_{-}$.
\par
The  temperatures of $\pi^{-}$ mesons have been
estimated  in the rapidity interval of\\
0.5$\leq$Y$\leq$2.1.
  A satisfactory fit for $\pi^{-}$ mesons
had been achieved by using a form involving two temperatures $T_{1}$ and $T_{2}$.
It was found that the QGSM underestimates $T_{2}$ by  $(10-15)\%$.
\par
The data have been analyzed  using the transverse momentum technique. The observed
dependence of the
$<P_{x}\hspace{0.01cm}^{\prime}(Y)>$ on Y shows the S-shape behaviour.
 The slope at midrapidity $F$ has been determined. The QGSM reproduces
the $<P_{x}>$ distribution satisfactorily, but underestimates
the parameter $F$.
Collective flow
\pagebreak
\begin{center}
\bf { 1.  INTRODUCTION }
\end{center}
\  \par
\par
Relativistic nucleus-nucleus collisions are very well suited for investigation
of the excited nuclear matter properties which
are the subject of intense studies both experimentally and theoretically.
 The  theoretical models predict
the formation of exotic states of nuclear matter.
\par
The experimental discovery of such states
is impossible without understanding the mechanism of collisions and studying
the characteristics of multiparticle production in nucleus-nucleus
interactions.
\par
The purpose of the present article is to study
properties  of $\pi^{-}$ mesons , produced in Mg-Mg collisions with the set-up
GIBS, which is the modified version of the set-up SKM-200 [1].
The choice of $\pi^{-}$ mesons is due to the fact, that they are dominantly
produced particles, they carry information about the dynamics of collisions and can be
unambiguously identified among other reaction products. It should be noted
also, that the production of $\pi^{-}$ mesons is the
predominant process at energies of the Dubna synchrophasotron.
\par
In previous articles [1,2] we studied the characteristics of $\pi^{-}$
mesons in collisions of various pairs of nuclei (He-Li, He-C, C-C, C-Ne, O-Ne,
Ne-Ne, C-Cu, C-Pb and O-Pb) and also the  results based on the part
of experimental data of Mg-Mg interactions (1390 interactions, 10414- $\pi^{-}$
mesons ) had been obtained. In paper [3] a good agreement of the CASIMIR
(CAScade Intranuclear Made In Rossendorf) cascade model calculations with
Mg-Mg interactions (6239 collisions - 50750 $\pi^{-}$ mesons)
experimental data
        on inclusive characteristics and correlations of secondaries
was shown. In this paper we have used the next generation
of intranuclear cascade models  -  the Quark Gluon String Model (QGSM) [4,5]
for comparison with experimental data. It is assumed that
interactions of identical nuclei (
Mg-Mg) give the possibility of better manifestation of nuclear effects than
the interactions of asymmetric pairs of nuclei.
\  \par
\newpage
\begin{center}
\bf { 2.EXPERIMENT }
\end{center}
\  \par
\par
The data were obtained using a 4$\pi$ spectrometer SKM-200 --- GIBS  of the
Dubna Joint Institute
for Nuclear Research. The facility  consists of a 2 m
streamer chamber with the fiducial volume 2$\times$1$\times$0.6 m$^{3}$, placed
in a magnetic field of $\sim$ 0.9 T and a triggering system.
The streamer chamber was exposed to beam of Mg nuclei accelerated in the synchrophasotron
up to a momentum of 4.3 GeV/c per incident nucleon.
The solid target
(Mg) in the form of a thin disc (with thickness 1.5 g/cm$^{2}$) was mounted inside
the chamber at a distance of 70 cm from the entrance window and at a height of
8 cm above the middle electrode.
The photographs of the events were taken using an optical system with 3 objectives.
The experimental set-up and the logic of the triggering system are presented in Figure 1.
The triggering system
allowed the selection of "inelastic" and
"central" collisions.
\par
The "inelastic" trigger, consisting of two sets of scintillation counters
mounted upstream (S$_{1}$ - S$_{4}$) and downstream (S$_{5}$, S$_{6}$) the chamber,
has been selecting all inelastic interactions of incident nuclei on a target.
\par
The "central" triggering system was consisting of the same upstream part as in the
"inelastic" system and of scintillation veto counters (S$_{ch}$, S$_{n}$),
registering a projectile and its charged and neutral spectator fragments,
in the downstream part. All counters were made from the plastic scintillators and
worked with photomultipliers PM-30. The S$_{1}$ counter with the scintillator
of 20$\times$20$\times$0.5 cm$^{3}$ size, worked in the amplitude regime and
identified the beam nuclei by its charge. The nuclei from the beam, going
to the target, have been selected using the profile counters S$_{2}$, S$_{3}$
with the plastic of 15 mm diameter and 3 mm thickness and "thin" counter
S$_{4}$ (15 mm and 0.1 mm correspondingly). The S$_{ch}$ counters (two
counters with plastic of 40$\times$40$\times$0.5 cm$^{3}$ size) were placed
at a distance of 4 m downstream from the target
 and registered the secondary charged particles, emitted from the target
within a cone of half angle $\Theta_{ch}$=$2.4^{0}$. The  S$_{n}$ counters
were registering the neutrons, emitted from the target in the same solid angle
$\Theta_{n}$=$2.4^{0}$. The S$_{n}$ telescope consisted of the five counters
of 40$\times$40$\times$2 cm$^{3}$ size, layered by 10 cm thick iron blocks.
 Thus, the trigger was selecting central collision events defined as those
without charged projectile spectator fragments and
spectator neutrons ( P/Z $>$ 3 GeV/c) emitted at
angles $\Theta_{ch}$=$\Theta_{n}$=$2.4^{0}$ ($\sim$ 4 msr) which corresponds
to a stripping nucleon transverse momentum of $\sim$ 180 MeV/c .
The trigger efficiency was  99 $\%$
and 80 $\%$  for charged and neutral projectile fragments, respectively.
  The trigger mode for each exposure is defined as
T($\Theta_{ch}$ ,$\Theta_{n}$) (where $\Theta_{ch}$ ,$\Theta_{n}$ are expressed
in degrees and rounded to the closest integer value). Thus Mg-Mg interactions
obtained on this set-up correspond to the trigger T(2,2).
The fraction
of such events is $\approx$ 4$\cdot$10$^{-4}$ among all inelastic interactions.
\par
Approximately 10$\%$ of $\pi^{-}$ mesons are lost mainly due to
absorption in the target and bad measurability of vertical tracks
which are localized
around a target rapidity value of -1.1 in the Mg-Mg rest frame. This leads to
a shift of mean $\pi^{-}$  rapidity by +0.1 relative to Mg-Mg c.m.s.
Average measurement errors of the momentum and production angle determination
for $\pi{-}$ mesons are $ \Delta P$$/$$P$ = 1.5$\%$,
$\Delta\theta=~ 0.3^{0}$.
\newpage
\begin{center}
\bf { 3.  QUARK GLUON STRING MODEL }
\end{center}
\  \par
\par
 In study of nucleus-nucleus collisions at high energy
several theoretical models
have been proposed [6]. The models allow one to test various
assumptions concerning the mechanism of particle production at extreme
conditions achieved only in nucleus-nucleus collisions. The models [7,8]
have taken into  consideration the decay of baryonic resonances. They
showed their
usefulness for the interpretation of the experimental data on pion
production. The comparison with the calculations of the Quark-Gluon String
Model (QGSM) can help us to understand properties of $\pi^{-}$ mesons,
produced in Mg-Mg collisions as well as to test  the validity of the
model in general. Below we briefly discuss the main points of the
meson production
mechanism in the framework of the QGSM.
\par
The model is presented in detail in papers [4,5]. The QGSM is based on
the Regge and string phenomenology of particle production in inelastic
binary hadron collision. To describe the evolution of the hadron and
quark-gluon phases, the model uses a coupled system of Boltzmann-like
kinetic equations. The nuclear collisions are treated as a mixture
of independent interactions of the projectile and target nucleons, stable
hadrons and short lived resonances. The QGSM includes low mass vector mesons
and baryons with spin 3/2, mostly  $\Delta$(3/2,3/2) via resonant
reactions. Pion absorption by NN quasi - deuteron pairs is also taken into
account. The coordinates of nucleons are generated according to a realistic
nuclear density. The sphere of the nucleus is filled with the nucleons at a
condition that the distance between them is greater than 0.8 fm. The nucleon
momenta are distributed in the range of 0 $\leq$ P $\leq$ $P_{F}$.
The maximum nucleon Fermi momentum is 
\par
\begin{center}
   $P_{F}$=$(3\pi^{-})^{2}h(\rho)^{1/3}$ \hspace{3cm}(1)
\end{center}
 where h=0.197 fm$\cdot$GeV/c, $\rho$(r) is nuclear density.
\   \par
\par
The procedure of event generation consists of 3 steps: the definition of
configurations on colliding nucleons, production of quark-gluon strings
and fragmentation of strings (breakup) into observed hadrons.
The model includes also the formation
time of hadrons. The QGSM has been extrapolated
to the range of intermediate energy (${\sqrt{s}}$ $\leq$ 4 GeV)
 to use it as a basic process
during the generation of hadron-hadron collisions. The masses of the 'strings'
produced at ${\sqrt{s}}$=3.6 GeV were small (usually not greater then 2 GeV),
 and they were fragmenting mainly
($\approx$90$\%$) through two-particle decays. For main NN and $\pi$N
interactions the following topological quark
diagrams [7] were used: binary, 'undeveloped' cylindrical, diffractive and planar.
The binary process makes a main contribution which is proportional to $1$/$P_{lab}$.
It corresponds to quark rearrangement without direct particle emission in
the string decay. This reaction predominantly results in the production of the
resonances (for instance, $ P + P$$\rightarrow$$ N + $$\Delta^{++}$), which are the main
source of pions. The comparable contributions to the inelastic cross section,
which however decreases with decreasing $P_{lab}$, come from the diagrams
corresponding to the "undeveloped" cylindrical diagrams and from the diffractive
processes.
The transverse momenta of pions produced in quark-gluon string
fragmentation processes are the
product of two factors: string motion on the whole as a result of transverse
motion of constituent quarks and $q$$\bar{q}$ production in string breakup. The
transverse motion of quarks inside hadrons was described by the Gaussian
distribution with variance $\sigma^{2}$ $\approx$ 0.3$(GeV/c)^{2}$. The
transverse momenta $k_{T}$ of produced $q$$\bar{q}$ pairs in the c.m.s. of the string
follow the dependence:
\  \par
\begin{center}
$W(k_{T})$=3B/$\pi(1+Bk^{2}_{T})^{4}$    \hspace{3cm}    (2)
\end{center}
where B=0.34 (GeV/c)$^{-2}$.
\par
The cross sections of hadron interactions were taken from the experiments.
Isotopic invariance and predictions of the additive quark model [9] (for
meson-meson cross sections, etc.) were used to avoid data deficiency.
 The resonance cross sections were assumed to be identical to the
stable particle cross sections with the same quark content.
For the resonances the tabulated widths were used.
\par
The QGSM simplifies the nuclear effects. In particular, coupling of nucleons
inside the nucleus is neglected, and the decay of excited recoil nuclear fragments
and coalescence of nucleons is not included.
\par
We have generated Mg-Mg  interactions using Monte-Carlo generator
COLLI, which is based on the QGSM.
The events have been traced through the detector and trigger filter.
\par
In the generator COLLI there are two possibilities to generate events:
1) at not fixed impact parameter $\tilde{b}$ and  2) at fixed $b$.
The events had been generated for $\tilde{b}$.
 The number of simulated events is
$\sim$ 4000.  From the impact parameter distribution
we obtained the mean value of
$<b>$=1.34 fm.
For the obtained value of $<b>$, we have generated
a total sample of 6200 events.  From the analysis of generated events
the pions with deep angles greater than 60$^{0}$ have been excluded, because
in the experiment the registration efficiency of
such vertical tracks is low.
\newpage
\begin{center}
\bf { 4.  KINEMATICAL CHARACTERISTICS OF PIONS }
\end{center}
\  \par
\par
We studied the kinematical characteristics of $\pi^{-}$ mesons, produced
in Mg-Mg collisions at a momentum of 4.3 GeV/c per nucleon, such as the
average multiplicity, momentum, transverse momentum, emission angle,
rapidity and corresponding distributions. In Table 1 the values of the
$\langle$$n_{-}$$\rangle$,
$\langle$$P$$\rangle$, $\langle$$P_{T}$$\rangle$, $\langle$$\Theta$$\rangle$,
$\langle$$Y$$\rangle$ for the experimental data and events generated
by the QGSM code for not fixed $\tilde{b}$ and $b$=1.34 fm
are listed. One can see, that the experimental and model values
coincide within the errors.
In Fig. 2-6 the corresponding experimental
and generated distributions of $n_{-}$, $P$, $\Theta$ , $P_{T}$ and
$Y$ are presented.
\par
The comparison of these  distributions allows
to conclude, that the QGSM satisfactorily describes the spectra.
The results obtained by the model in the two regimes
are consistent and it seems, that in our experiment the value of
b=1.34 fm for Mg-Mg is most probable.
\par
We used two Lorenz invariant variables to describe the main features of the
$\pi^{-}$ mesons, produced in nucleus-nucleus collisions: the rapidity Y and
the transverse momentum P$_{T}$. For this purpose we investigated rapidity
distributions in various regions of P$_{T}$: P$_{T}$ $\leq$ 0.2,
0.2 $\leq$ P$_{T}$$<$ 0.3,
0.3 $\leq$ P$_{T}$$<$ 0.5, P$_{T}$ $\geq$ 0.5. The distributions
(Fig.7) have the characteristic Gaussian form.
The form of $Y$
distributions changes with increase of the transverse momentum
of $\pi^{-}$ mesons: the fraction of pions increases in the central
region and decreases in fragmentational regions of colliding nuclei.
The $Y$ distributions are narrowing. The dispersion is changing
from $D_{Y}$=1.16 for
P$_{T}$ $\leq$ 0.2 to  $D_{Y}$=0.70 for P$_{T}$ $\geq$ 0.5.
The analysis of the $Y$ distributions shows, that the central regions
of these distributions are enriched with pions of large transverse
momentum (as compared to fragmentation regions of the colliding nuclei):
$\langle$$P_{T}$$\rangle$ =$0.184\pm0.004$ GeV/c  for  Y$<$ 0.2,
$\langle$$P_{T}$$\rangle$ =$0.247\pm0.002$ GeV/c  for
0.7$\leq$Y$\leq$1.6,
$\langle$$P_{T}$$\rangle$ =$0.189\pm0.003$ MeV/c  for  Y$>$ 2.
The QGSM reproduces the $Y$ distributions in the various regions
of  P$_{T}$ well enough (Fig.7). The corresponding mean values of $Y$ for experimental
and model data are listed in Table 2.
\par
Practically all theoretical models are based on the dependence of the average
kinematical characteristics on the impact parameter {\it{b}}. As {\it{b}} is
experimentally unmeasurable, the estimation of {\it{b}} may be  obtained
on the basis of the number
of nucleons $\nu_{p}$ of the projectile participating in the interaction,
which in turn is
correlated with the multiplicity of observed $\pi^{-}$ mesons n$_{-}$.
Therefore the study of dependence of a given variable (P$_{T}$, etc)
on {\it{b}} can be qualitatively replaced by the study of the dependence
of the same variable on $n_{-}$.
\par
In Fig.8 the dependence of $<P_{T}>$ on $n_{-}$ is presented.
The dashed line corresponds to data, derived from the experiments on N-N
collisions [10] at our energy (the values are averaged over all n$_{-}$ values).
 One
can see, that $<P_{T}>$ slightly
decreases with multiplicity.
In the model of independent collisions [11], nucleus-nucleus interactions are
considered as a superposition of independent nucleon-nucleon collisions. In this
case, the coincidence with the N-N collisions is expected.
Fig.8 shows, that the QGSM reproduces
these dependences well. The reason caused some quantitative disagreement in
$<P_{T}>$
may be attributed to the  nuclear effects, which are essential
in Mg-Mg collisions.
\newpage
\begin{center}
\bf { 5.  THE TEMPERATURE  OF {\large{$\pi^{-}$}} MESONS }
\end{center}
\  \par
\par
An excited system of hadrons is characterized by its temperature. The $\pi^{-}$
mesons
temperature  was estimated by means
of: 1. spectra of inclusive kinetic energy E$_{K}$ and 2. spectra of
transverse momentum P$_{T}$.
\par
Non-invariant inclusive spectra d$^{3}\sigma/d\vec{P}=(E^{*}P^{*})^{-1}dN/
dE^{*}_{K}$
(P$^{*}$ is the momentum, E$^{*}$- the total energy and E$^{*}_{K}$ the kinetic
energy of the particle in the c.m.s.) have been analyzed.
The experimental spectrum was fitted by a simple exponential law:
\begin{center}
$F(E^{*}_{K})$=$(E^{*}P^{*})^{-1}dN/dE^{*}_{K}$=
A$_{1}$$\cdot exp (-E^{*}_{K}/T_{1})$ +
A$_{2}$$\cdot exp (-E^{*}_{K}/T_{2})$\hspace{2cm}(3)
\end{center}
\par
$T$ is related to the average kinetic energy of a given type of particles and
thus characterizes the nuclear matter temperature at the expansion stage when
such particles are emitted. Therefore, the parameter $T$ is usually called an
average or inclusive temperature.
The temperature may also be deduced from transverse momentum distributions.
This method was suggested by Hagedorn in the framework of the
thermodynamic model [12,13].
This model assumes that in the collision process the
hot sources (one or several) are
formed and move together along collision axis. Their longitudinal velocity
and temperature obey energy-momentum conservation laws.  Some authors [12-14]
state that transverse momentum distributions are preferable because of their
Lorentz-invariance in respect to longitudinal boosts. Transverse
momentum distributions
were described by [12-15]:
\  \par
\begin{center}
$dN/dP_{T}$
$\approx$ const $P_{T}$$\cdot$$E_{T}$ $\cdot$$( exp(- E_{T}/T_{1})$ +
$( exp(- E_{T}/T_{2}))$\hspace{2cm}(4)
\end{center}
\begin{center}
$E_{T}$=$(P^{2}_{T}+m^{2})^{1/2}$
\end{center}
\par
 The pion spectra
can be well fitted by a sum of two exponentials (two
temperatures $T_{1}$ and $T_{2}$ ).
Experimental and generated spectra of the non-invariant
kinetic energy $E_{kin}$
and P$_{T}$ have been approximated using formulas
(3) and (4) in the
rapidity interval 0.5$\leq$Y$\leq$2.1, which corresponds to the pionization
region.
The fitted parameters are presented in the
Table 3.
\par
We can see from the Table, that within errors parametrizations
 (3) and (4) give consistent results for
$T_{1}$ and $T_{2}$. Calculated temperatures are consistent
within both approaches, but the model underestimates high temperature
component by about $(10-15)\%$.
\par
At our energy Hagedorn model [13] predicts only one temperature
$T_{\pi^{-}}$=(115 - 120) MeV, which agrees with our value of $T_{2}$.
 Several possible mechanisms
have been proposed to explain two temperatures [7,16,17,18]. In [16] the
presence of two temperatures for pions is explained by two
mechanisms of pion production: direct ($T_{2}$) and via $\Delta$ resonance
decay ($T_{1}$).  In [17] the necessity of two temperatures is found to be
caused by the different contributions of the $\Delta$
resonance produced during the early and the late stages of the nucleus-nucleus
collisions, and due to the energy dependence of the pion and $\Delta$
absorption cross sections. In paper [18] this effect was quantitatively explained
by taking into account the finiteness of the number of particles in the statistical
ensemble and the resonance absorption mechanisms.
\  \par
\ \par
\begin{center}

     {\bf{ 6. TRANSVERSE MOMENTUM ANALYSIS TECHNIQUE}}
\end{center}
\   \par
\par
The extraction of properties of the hot and dense hadronic matter from pion
observables critically depends on our understanding of the pion-production
dynamics. Hydrodynamical calculations predicted the formation of compression
waves in nuclear matter in high energy heavy ion collisions [19]. Using
the transverse momentum  analysis technique developed by Danielewicz
and Odyniec [20], nuclear collective flow has already been observed
for protons, light nuclei and $\Lambda$ - hyperons emitted in
nucleus-nucleus
collisions at energies 0.4$\div$1.8 GeV/nucleon at the BEVALAC, GSI-SIS [21-24],
and at 11$\div$14 GeV/nucleon at AGS [25,26]. Due to the small mass
of pions compared to that of baryons, it has been pointed out that the pions
might serve as a good probe of any hydrodynamical flow [27]. Moreover, as pions
are mainly coming from the decay  of $\Delta$ resonances in the relativistic
nucleus-nucleus interactions, the remnant of the collective flow carried by
$\Delta$ resonances might be seen in the distributions of the
final state pions. In search for
collective flow signatures among final-state pions the transverse momentum
 technique has also been applied to pions from relativistic nuclear
reactions [21,23]. Looking for flow effects
in pion emission, they have [21,23] measured the correlation between the
reaction plane, deduced from a transverse-momentum analysis performed
on light baryons (protons), and the momenta of charged pions emitted in
collisions.
\par
 P.Danielewicz and G.Odyniec have proposed
an exclusive way to analyse the momentum contained in direct sidewards
emission and presented the data in terms of the mean transverse momentum
per nucleon in the reaction plane $<P_{x}(Y)>$ as a function of the rapidity.
The reaction plane is determined by the vector
$\overrightarrow{Q_{j}}$=$\sum\limits_{i\not=j}\limits^{n}$$\omega_{i}$
$\overrightarrow{P_{{\perp}i}}$ \hspace{6cm} and the incident beam direction.
Here $P_{{\perp}i}$ is the transverse momentum  of particle $i$, and $n$ is
the number of particles in the event. Pions are not included.
The weight $\omega_{i}$
is taken as 1 for y$_{i}$$>$ y$_{cm}$ and -1 for y$_{i}$$<$ y$_{cm}$,
where y$_{cm}$ is c.m.s. rapidity and y$_{i}$ is the rapidity of
particle $i$.
 The transverse momentum of each
particle  in the estimated reaction plane is calculated as
$P_{xj}\hspace{0.01cm}^{\prime}$ = $\{ \overrightarrow{{Q_{j}}}\cdot
\overrightarrow{P_{{\perp}j}}$ $/$
$\vert\overrightarrow{{{Q_{j}}}}\vert\} $.
\par
 The average transverse momentum $<P_{x}\hspace{0.01cm}^{\prime}(Y)>$ is
obtained by averaging over all
events in the corresponding intervals of rapidity.
\par
 For Mg-Mg collisions we have inclusive data - only
$\pi^{-}$ mesons are measured.
As the
$\pi^{-}$ mesons are emitting mainly from decays of $\Delta$ isobars
(at least $\sim$ 80$\%$) [28],
we decided to investigate whether a single  $\pi^{-}$ meson
of the event knows something about its origin and the question
whether they are collectively correlated. For this purpose we applied the
technique of P.Danielewicz and G.Odyniec to our data.
 As we have inclusive data, we constructed $\overrightarrow{Q}$ vector
from the $\overrightarrow{P_{{\perp}i}}$ of only $\pi^{-}$ mesons
event by event for the events
with multiplicity $n_{-} > 7$.
\par
It is known [20], that the estimated reaction plane
differs from the true  one,
due to the finite number of particles in each event.
The component $ P_{x}$ in the true reaction plane is systematically larger
then the component $P_{x}\hspace{0.01cm}^{\prime}$
in the estimated plane,   hence\\
$<P_{x}>$=$<P_{x}\hspace{0.01cm}^{\prime}>/<cos\varphi>$
where $\varphi$ is the angle between the estimated and true planes.
The  correction factor
$K$=1 $/$ $< cos\varphi >$ is subject to a large uncertainty,
especially for low multiplicity.
According to [20], for the definition of $< cos\varphi >$ we divided
each event randomly into two equal sub-events,
constructed vectors $\overrightarrow{Q_{1}}$ and
$\overrightarrow{Q_{2}}$ and estimated azimuthal angle  $\varphi_{1,2}$
between these two vectors. $<cos\varphi>$=$<cos(\varphi_{1,2}/2)>$.
We defined the correction factor $K$, averaged
over all the multiplicities. The value of $K$ has been obtained
 $K$=1.51$\pm$0.05.
\par
Fig. 9 shows the
dependence of the estimated $<P_{x}\hspace{0.01cm}^{\prime}(Y)>$
on Y for pions. The data exhibit $S$- shape behaviour similar to
the form of the $<P_{x}>$ spectra for protons [21,22,24] and pions [21,23]
obtained
at lower energies and identified as nucleon and pion collective flow.
The slope at midrapidity has been extracted from a linear fit to the data
for Y between 0.2$\div$2. The straight line in Fig.9 shows the result of
this fit. The value of $F$ is --  $F=48\pm5$  (MeV/c).
$F$ is normally lower than the true value because
$P_{x}\hspace{0.01cm}^{\prime} < P_{x}$, thus
 the obtained value of the parameter $F$ can be considered
as lower limit.
In the model calculations the reaction plane is known a priori and is refered
as the true reaction plane.
 For simulated events
the component in the true reaction plane $ P_{x}$ had been calculated.
The dependences of  $<P_{x}(Y)>$ on $Y$ (for both $\tilde{b}$ and $b$) are
shown in Fig.9. The experimental and QGSM results coincide
within the errors.
For the visual presentation,
we approximated these dependences by polynoms (the curves in Fig.9).
From the comparison of the dependences  of $<P_{x}(Y)>$ on $Y$ obtained
by the model in two regimes - for $\tilde{b}$ and $b$, one can conclude,
that the results are consistent. The values of $F$, obtained from the
QGSM are: $F=53\pm3$ (MeV/c) --- for not fixed $\tilde{b}$;
$F=51\pm4$ (MeV/c) --- for $b$=1.34. One can see, that the QGSM underestimates
this parameter, as the value of $F$ from the experimental data is lower
limit, because it is not multiplied on the uncertainty factor $K$.
\par
The origin of the particular
shape of the $\overrightarrow{P_{x}}$ spectra for pions had been studied
in [28-30].
The investigation revealed, that the effect of baryon collective flow
on the pion transverse momentum distribution is negligible and the origin of
the  in-plane transverse momentum of pions is not
the remnant of the $\Delta$ flow [29], but
the pion scattering process (multiple $\pi N $ scattering) [28]
and the pion absorption [29,30]. The $<P_{x}>$ distribution of pions
is not a collective effect in the sense of the nucleonic bounce-off: the
observable is the same, but the cause is different.
\par
In the framework of the Isospin Quantum Molecular Dynamics (IQMD) model
[28] and the hadronic BUU transport model [30] it had been obtained, that
in central Au-Au collisions ($b \leq 3~ fm$) at E=1--2 GeV/n and
La-La collisions at E=0.8 GeV/n the pion average transverse momentum $<P_{x}>$
have the same sign as that of nucleons in both forward and backward
rapidities. The correlation of nucleon and pion flow
 had been observed
experimentally by the DIOGENE group for central Ne-NaF, Ne-Nb and Ne-Pb collisions
($b \leq 3~ fm$) at E=0.8 GeV/n [23]. These results are in agreement with our findings.
With increase of the impact parameter $b$, the $<P_{x}>$  of pions changes
its sign [28,30]. Therefore as reaction goes from central to semicentral and
peripheral ($b \geq 4~ fm$) the $<P_{x}>$ distribution of pions undergoes
a transition from being correlated to anticorrelated with that of nucleons.
The anticorrelation of nucleons and pions in [28] was explained as due
to multiple $\pi ~N$ scattering. However in [30] it had been shown, that
the anticorrelation is a manifestion of the nuclear shadowing effect
of the target- and projectile-spectator  through both pion rescattering and
reabsorptions.
\newpage
\begin{center}
\bf { 7.  CONCLUSIONS }
\end{center}
\  \par
\par
A study of pion production in central Mg-Mg collisions was carried out.
\par
1. The average kinematical characteristics of pions such as multiplicity, momentum,
transverse momentum, emission angle, rapidity
 and corresponding distributions have been obtained. The QGSM satisfactorily
describes the experimental results.
\par
2.
It has been shown, that  $<P_{T}>$ slightly
decreases with multiplicity. This average kinematical characteristic
are similar to the characteristics of N-N collisions at the same energy.
 The reason of some quantitative disagreement in
$<P_{T}>$
may be attributed to the  nuclear effects, which are essential
in Mg-Mg collisions. The QGSM reproduces the dependence of
$<P_{T}>$ on $n_{-}$.
\par
3. The temperature of $\pi^{-}$ mesons has been determined from $E_{K}$ and
$P_{T}$ spectra in the rapidity interval 0.5$\leq$Y$\leq$2.1.
The sum of two exponentials is necessary
to reproduce the data.
The temperatures have been obtained from 
spectra generated by QGSM. It underestimates $T_{2}$ about (10-15) $\%$.
\par
4. The data have been analyzed event by event using transverse momentum
technique.
The results are presented
in terms of the mean transverse momentum
$<P_{x}\hspace{0.01cm}^{\prime}(Y)>$ as a function of the Y.
The observed dependence of the
$<P_{x}\hspace{0.01cm}^{\prime}(Y)>$ on Y shows the S-shape behaviour.
 The slope at midrapidity $F$ had been extracted. The QGSM reproduces
the $<P_{x}>$ distribution, but underestimates the parameter $F$.
\   \par
\   \par
ACKNOWLEDGEMENTS
\   \par
\   \par
\par
We would like to thank professor N.Amaglobeli for  his  continuous
support. We  are  indebted  to  M.Anikina , A.Golokhvastov, S.Khorozov,
J.Lukstins, \fbox{L.Okhrimenko} for helping in obtaining the data and
many valuable  discussions.
We are very grateful to N.Amelin for providing us with the QGSM
code program COLLI, also to
Z. Menteshashvili for his continuous support during
the preparation of the article and reading the manuscript.
\par
The research described in this publication was made possible in part
by Grant  MXP000 from the International Science Foundation and  Joint Grant
MXP200 from the International Science Foundation and Government of Republic
of Georgia.
\newpage
\par
\par
\newpage
\begin{center}
\bf{ FIGURE CAPTIONS  }
\end{center}
{\bf Fig.1.}
Experimental set-up. The trigger and the trigger distances are not to
scale. \\
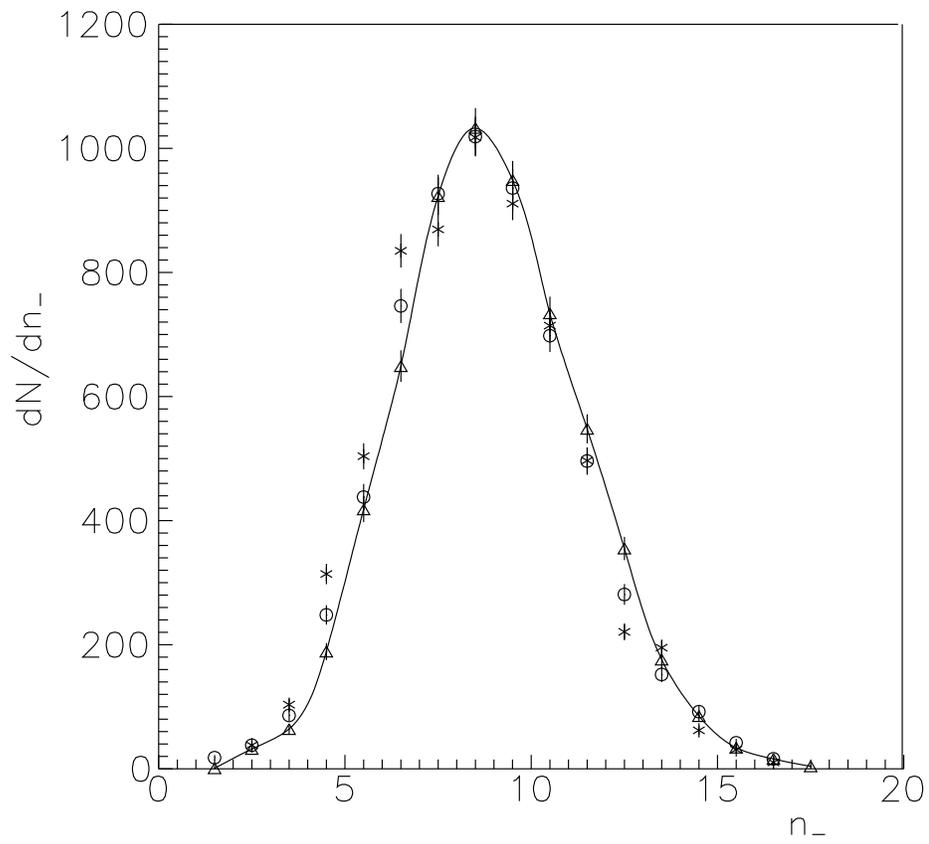
{\bf Fig.2.}
The multiplicity
distribution of $\pi^{-}$ mesons. $\circ$ -- the experimental data,
curve --  QGSM generated data for $b$=1.34 fm,
$\ast$ -- QGSM generated data for $\tilde{b}$.\\
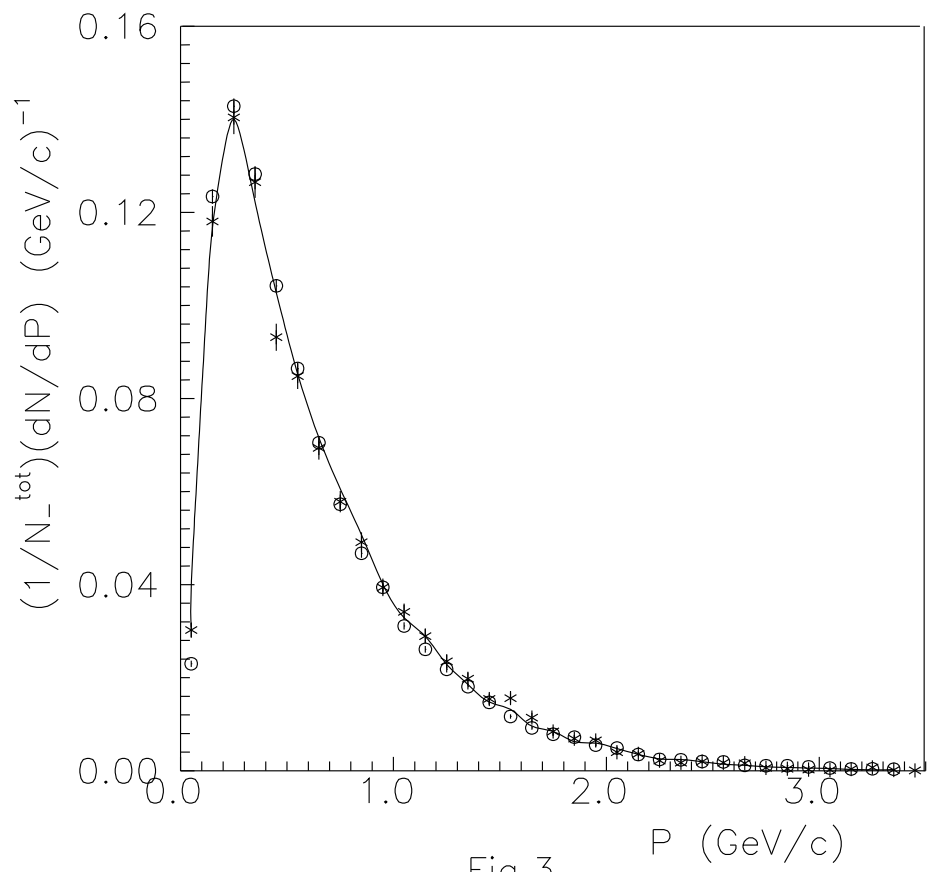
{\bf Fig.3.}
The momentum
distribution of $\pi^{-}$ mesons. $\circ$ -- the experimental data,
curve --  QGSM generated data for $b$=1.34 fm,
$\ast$ -- QGSM generated data for $\tilde{b}$.\\
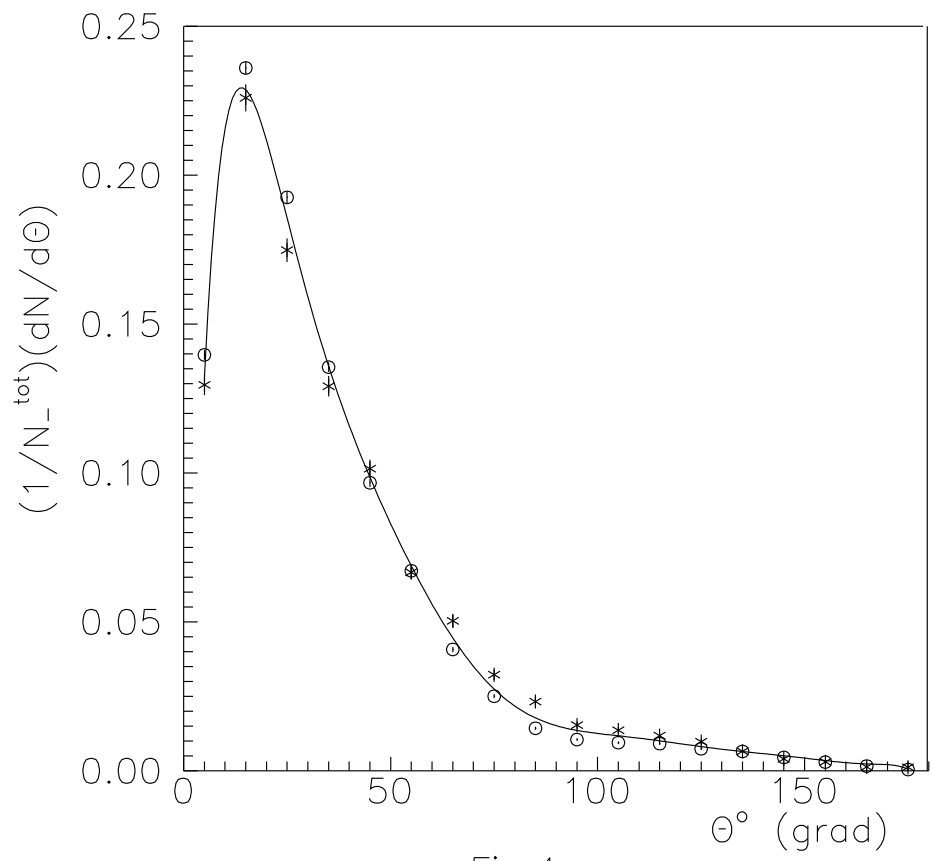
{\bf Fig.4.}
The emmision angle
distribution of $\pi^{-}$ mesons. $\circ$ -- the experimental data,
curve --  QGSM generated data for $b$=1.34 fm,
$\ast$ -- QGSM generated data for $\tilde{b}$.\\
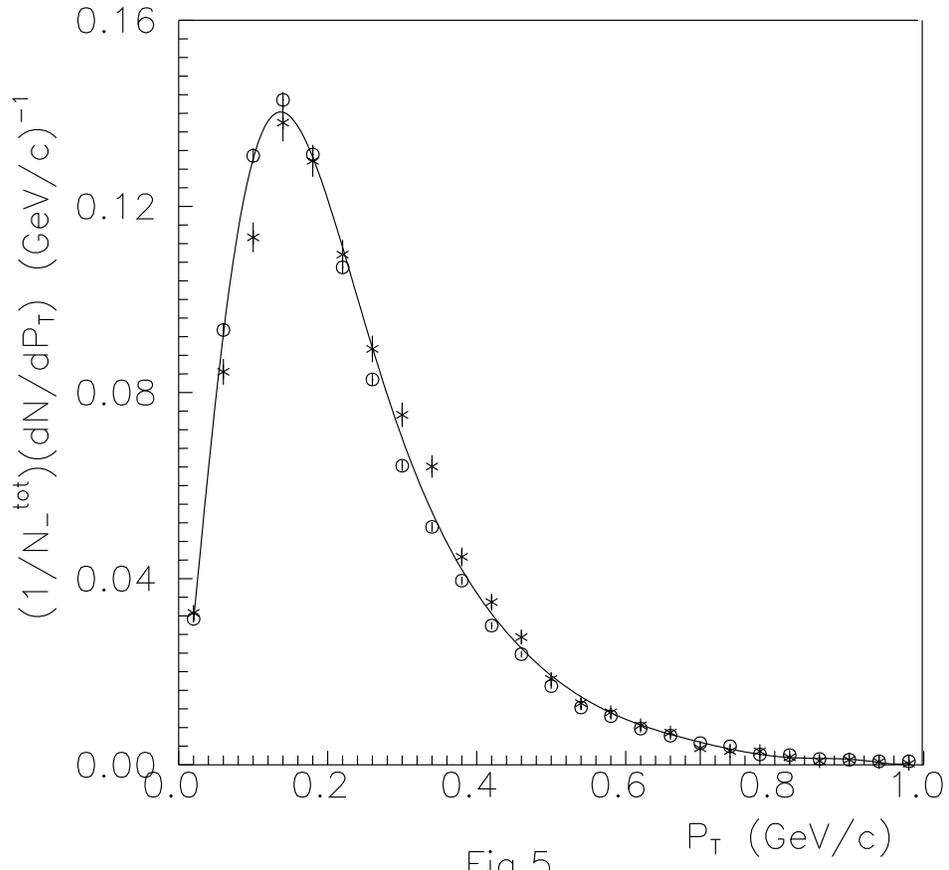
{\bf Fig.5.}
The  transverse momentum
distribution of $\pi^{-}$ mesons. $\circ$ -- the experimental data,
curve --  QGSM generated data for $b$=1.34 fm,
$\ast$ -- QGSM generated data for $\tilde{b}$.\\
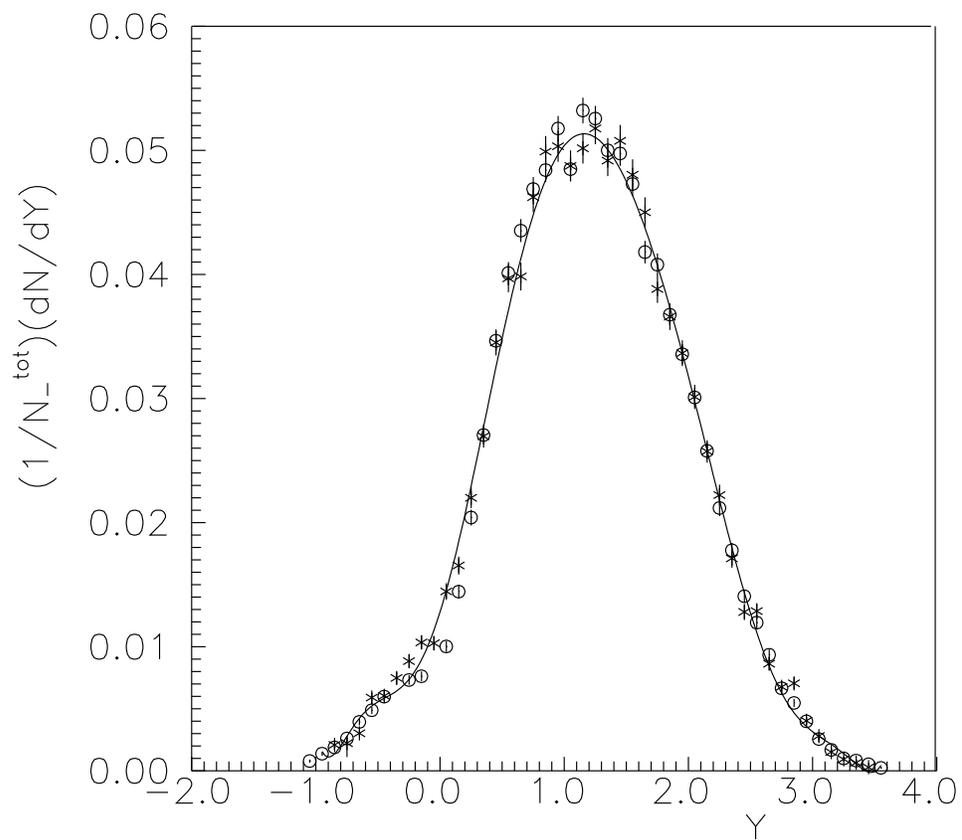
{\bf Fig.6.}
The  rapidity
distribution of $\pi^{-}$ mesons. $\circ$ -- the experimental data,
curve --  QGSM generated data for $b$=1.34 fm,
$\ast$ -- QGSM generated data for $\tilde{b}$.\\
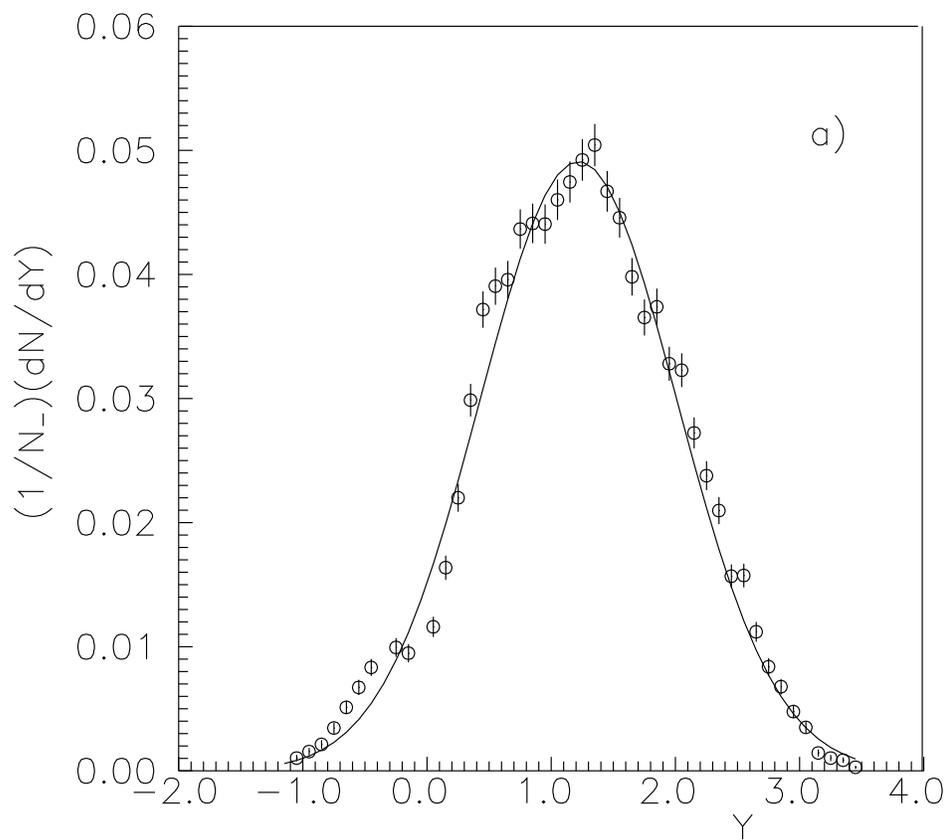
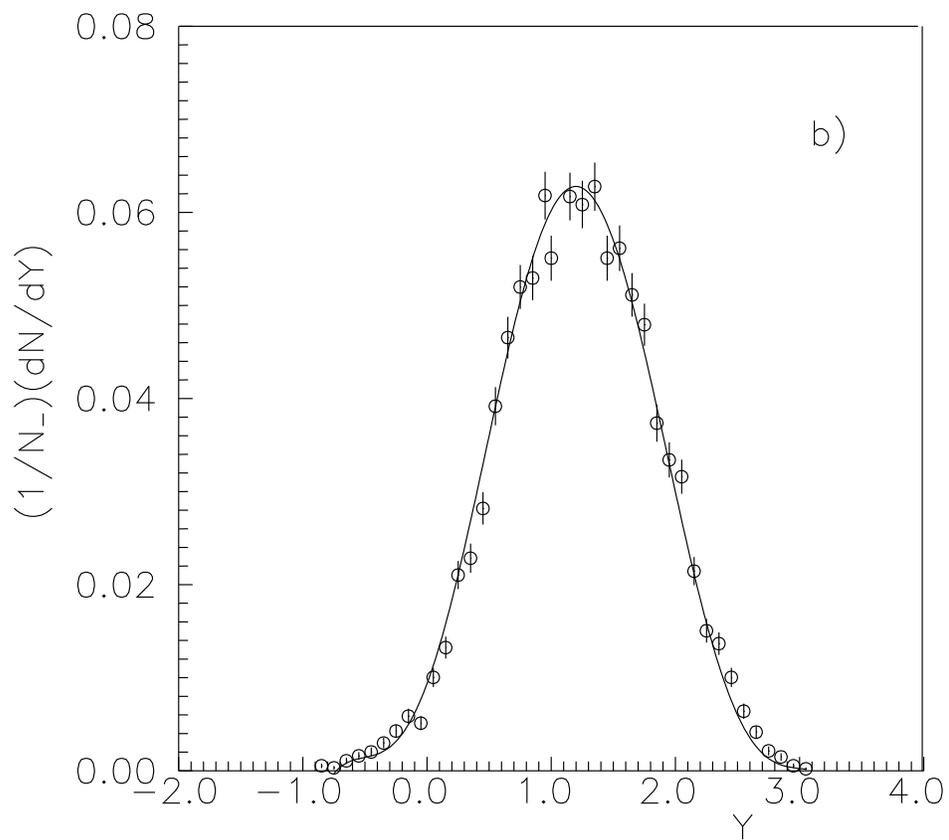
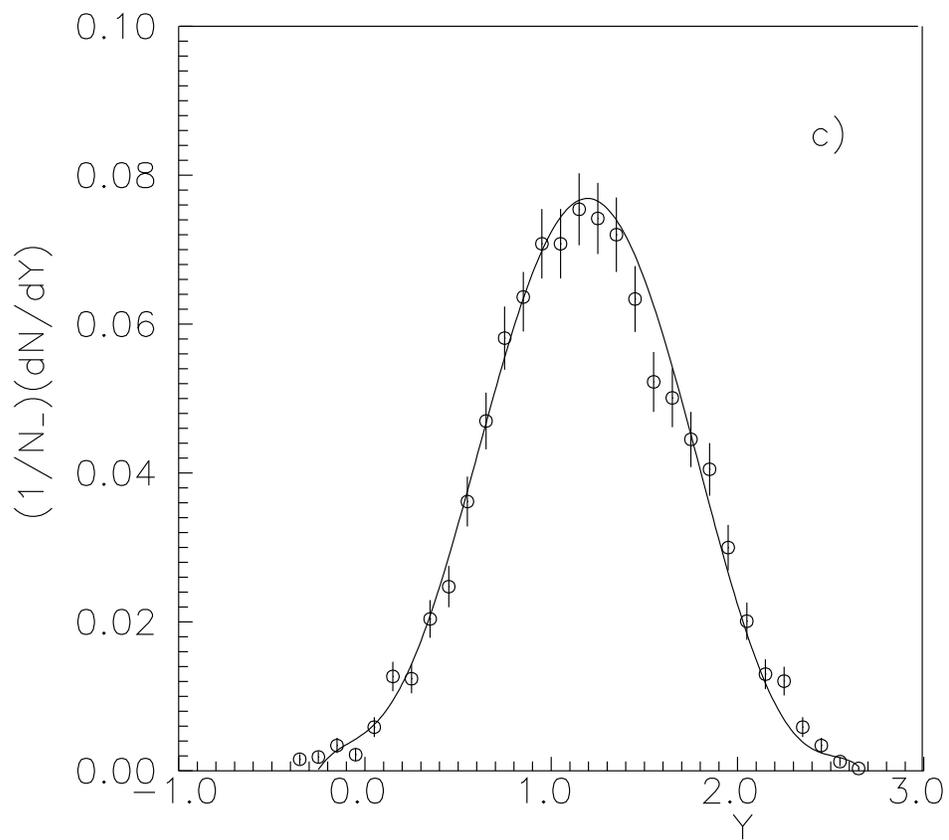
{\bf Fig.7.}
The  rapidity
distribution of $\pi^{-}$ mesons in various intervals of $P_{T}$.
a) $P_{T}$$<0.2$ ;   $\circ$ -- the experimental data,
curve --  QGSM generated data for $b$=1.34 fm.\\
b) $0.2\leq$ $P_{T}$$<0.3$ ;   $\circ$ -- the experimental data,
curve --  QGSM generated data for $b$=1.34 fm.\\
c) $P_{T}$$>0.5$ ;   $\circ$ -- the experimental data,
curve --  QGSM generated data for $b$=1.34 fm.\\
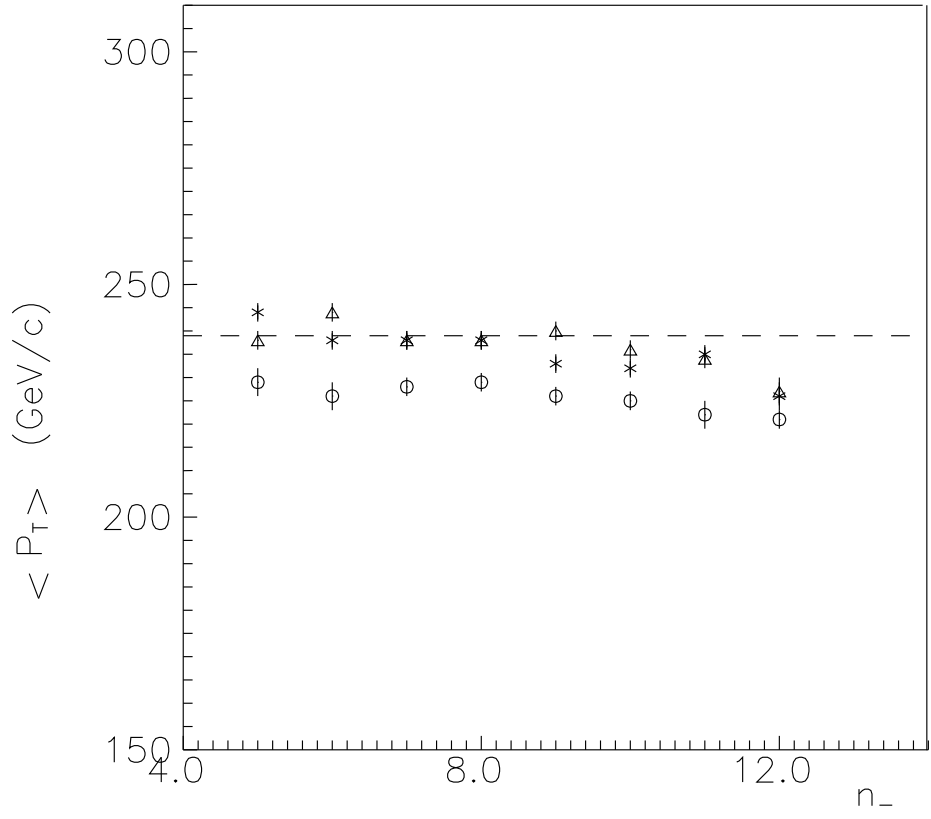
{\bf Fig.8.}
 The dependence of the average transverse momentum $<P_{T}>$ on the
$n_{-}$ variable. $\circ$ -- the experimental data,
$\bigtriangleup$ -- QGSM generated data for $b$=1.34 fm,
$\ast$ -- QGSM generated data for $\tilde{b}$.
Dashed line shows the $<P_{T}>$ value for N-N
collisions at 4.3 GeV/c.\\
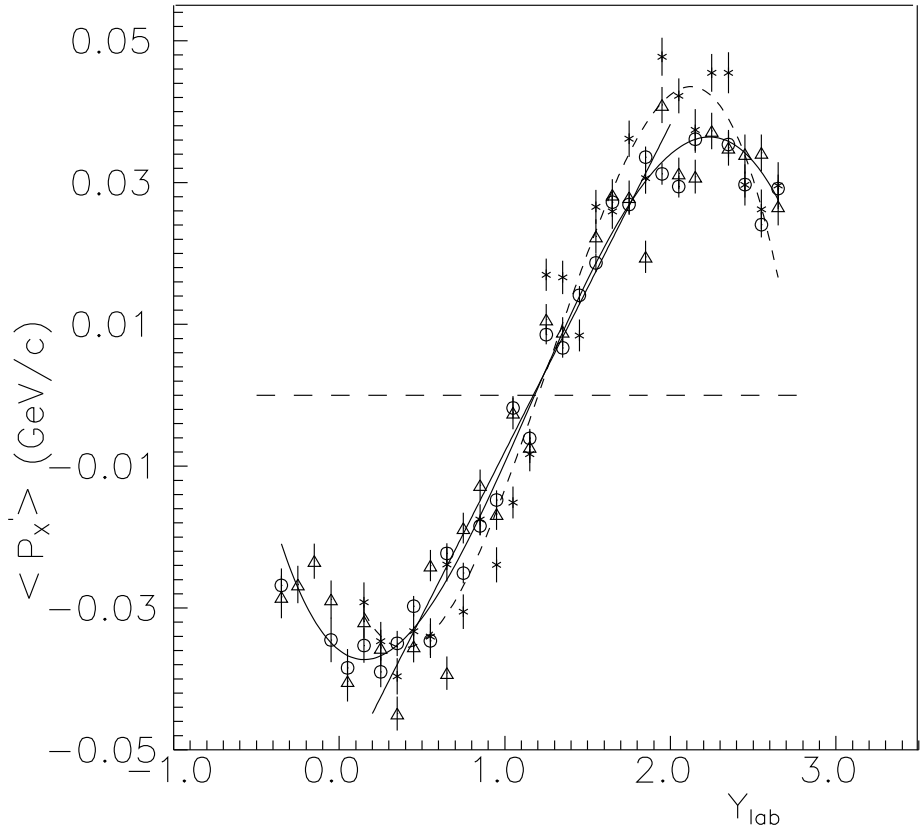
{\bf{Fig.9}}
The dependence of $< P_{x}\hspace{0.01cm}^{\prime}(Y) >$ on Y$_{Lab}$.
 $\circ$ -- the experimental data,
$\bigtriangleup$ -- the QGSM generated data for fixed $b$= 1.34 fm,
$\ast$ -- the QGSM generated data for $\tilde{b}$.
The solid
line is the result of the the linear approximation of experimental data
 in the interval of
 Y - 0.2 $\div$ 2.0
. The curves for
visual presentation of the QGSM events (solid - for fixed $b$,
 dashed -for $\tilde{b}$)
- result of approximation by 4-th order polynomial function.
\newpage
\begin{center}
\bf{ TABLE CAPTION  }
\end{center}
{\bf{Table.1}} The average kinematical characteristics of
$\pi^{-}$ mesons.\\
{\bf{Table.2}} The $\langle$$Y$$\rangle$ and dispersion $D_{Y}$
for various intervals of $P_{T}$.\\
{\bf{Table.3}} The
temperatures of $\pi^{-}$ mesons - result of approximation by (3) and (4).\\
\newpage
Table.1  The average kinematical characteristics of
$\pi^{-}$ mesons.\\
\   \par
\   \par
\   \par
\   \par
\begin{tabular}{|c|c|c|c|c|c|}
\hline
 & $\langle$$n_{-}$$\rangle$ &
   $\langle$$P$$\rangle$ $(GeV/c)$ & $\langle$$\Theta$$\rangle$
&$\langle$$P_{T}$$\rangle$$(GeV/c)$
& $\langle$$Y$$\rangle$\\
 \hline
& &&&&\\
experiment&  8.2$\pm$0.1 &
  0.628$\pm$0.003 & 33.8$^{0}$$\pm$0.1$^{0}$ &0.229$\pm$0.003 &
1.22$\pm$0.03 \\
& &&&&\\
\hline
& &&&&\\
QGSM& &&&&\\
$b$=1.34 fm  &  8.3$\pm$0.1 &
  0.632$\pm$0.003 & 35.5$^{0}$$\pm$0.1$^{0}$ &0.237$\pm$0.003 &
 1.20$\pm$0.02   \\
& &&&&\\
\hline
& &&&&\\
QGSM& &&&&\\
$\tilde{b}$  &  8.1$\pm$0.2 &
  0.629$\pm$0.005 & 36.5$^{0}$$\pm$0.3$^{0}$ &0.238$\pm$0.004 &
 1.18$\pm$0.03   \\
& &&&&\\
\hline
\end{tabular}
\newpage
Table.2 The $\langle$$Y$$\rangle$ and dispersion $D_{Y}$
for various intervals of $P_{T}$.
\   \par
\   \par
\   \par
\   \par
\   \par
\   \par
\begin{tabular}{|c|c|c|c|c|c|}
\hline
  & &$P_{T}$$<$0.2 & 0.2$\leq$$P_{T}$$<$0.3 &
 0.3$\leq$$P_{T}$$<$0.5 & $P_{T}$$\geq$0.5 \\  \hline
&&&&&  \\
experiment& $\langle$$Y$$\rangle$ & 1.23$\pm$0.01 & 1.22$\pm$0.02 &
1.22$\pm$0.02 & 1.20$\pm$0.02  \\
    &$D_{Y}$ & 1.16$\pm$0.02 & 0.95$\pm$0.01 & 0.85$\pm$0.02 &
   0.70$\pm$0.01  \\
&&&&&  \\
\hline
&&&&&  \\
  QGSM & & & & &   \\
$b$=1.34 fm &$\langle$$Y$$\rangle$ & 1.20$\pm$0.02 & 1.21$\pm$0.02
& 1.19$\pm$0.02&1.18$\pm$0.02  \\
   &$D_{Y}$  &1.24$\pm$0.01 &0.97$\pm$0.01 &0.80$\pm$0.01&
 0.70$\pm$0.02  \\
&&&&&  \\
\hline
\end{tabular}
\newpage
Table.3 The
temperatures of $\pi^{-}$ mesons - result of approximation by (4) and (5).
\  \par
\  \par
\  \par
\begin{tabular}{|c|c|c|c|}
\hline
&    &    &    \\
\multicolumn{1}{|c|}{$A_{P}$ - $A_{T}$}& \multicolumn{1}{c|}{Selection}&
 \multicolumn{1}{c|}{$T_{1}$ (MeV)
$T_{2}$ (MeV)} & \multicolumn{1}{c|}{$T_{1}$ (MeV) $T_{2}$ (MeV)}  \\

    & criteria & using (1) &  using(2)      \\
\hline
    &    &       & \\
experiment  & 0.5$\leq$Y$\leq$2.1& 55$\pm$1   113$\pm$2   & 61$\pm$1    110$\pm$3
 \\
              &    &       & \\
    &    &       & \\
\hline
    &    &       & \\
  QGSM & 0.5$\leq$Y$\leq$2.1& 52$\pm$8   103$\pm$1   & 66$\pm$7    95$\pm$5
\\
$b$=1.34 fm            &    &   & \\
              &    &       & \\
\hline
\end{tabular}

\begin{thebibliography}{99}
\bibitem{}{ M.X. Anikina {\it{et al}}: {\it{Phys.Rev.}} C{\bf{33}} (1996) 895}
\bibitem{}{L.V. Chkhaidze {\it{et al}}: {\it{Z.Phys.}} C{\bf{54}} (1992) 179}
\bibitem{}{L.V. Chkhaidze {\it{et al}}: {\it{J.Phys.}} G{\bf{22}} (1996) 641}
\bibitem{}{N. Amelin {\it{et al}}: {\it{Yad.Fiz.}} {\bf{52}} (1990) 272}
\bibitem{}{N. Amelin {\it{et al}}: {\it{Phys.Rev.}} C{\bf{44}} (1991) 1541}
\bibitem{}{S. Nagamiya and M. Gyulassy: {\it{Adv.in.Nucl.Phys.}} {\bf{13}} (1984) 201
 and references therein}
\bibitem{}{B.A. Li and W. Bauer: {\it{Phys.Rev.}} C{\bf{44}} (1991) 450}
\bibitem{}{M.D. Zubkov: {\it{Yad.Fiz.(Soviet Journal of Nuclear
 Physics)}} {\bf{55}} (1989) 455}
\bibitem{}{V.V. Anisovich {\it{et al}}: {\it{Nucl. Phys.}} B{\bf{133}} (1978) 477}
\bibitem{}{K. Beshliu {\it{et al}}: {\it{Yad.Fiz.}} {\bf{43}} (1986) 808}
\bibitem{}{S.A. Khorozov: {\it{JINR  Dubna Report}} 2-80142 (1980)}
\bibitem{}{R. Hagedorn: {\it{CERN  Geneva Preprint}} TH{\bf-{3684}}
 (1984) 59}
\bibitem{}{R. Hagedorn: {\it{Phys.Lett.}} B{\bf{97}} (1980) 136}
\bibitem{}{V. Gudima {\it{et al}}: {\it{Phys.Elem.Part.
 At.Nucl.}} {\bf{17}} (1986) 1093}
\bibitem{}{R. Stock: {\it{Phys.Rep.}} {\bf{135}} (1986) 261}
\bibitem{}{S. Nagamiya: {\it{Phys.Rev.Lett.}} {\bf{49}} (1982) 1383}
\bibitem{}{D. Hahn and N. Glendenning: {\it{Phys.Rev.}}
 C{\bf{37}} (1988) 1053}
\bibitem{}{M.D. Zubkov: {\it{Yad.Fiz.}} {\bf{49}} (1989) 1751}
\bibitem{}{H. Stocker and W. Greiner: {\it{Phys.Rep.}}
 {\bf{137}}  (1986)  277 and references therein}
\bibitem{}{P. Danielewicz and G. Odyniec: {\it{Phys.Lett.}}
 B{\bf{157}} (1985) 146}
\bibitem{}{P. Danielewicz {\it{et al}}: {\it{Phys.Rev.}}
 C{\bf{38}} (1988) 120}
\bibitem{}{O. Beavis {\it{et al}}: {\it{Phys.Rev.}} C{\bf{45}} (1992) 299}
\bibitem{}{J. Gosset {\it{et al}}: {\it{Saclay Rapport DPh-N/Saclay}};
                    {\bf{2469B}} (1987) 5}\\
          {J. Gosset {\it{et al}}: {\it{Phys.Rev.Lett.}} {\bf{62}} (1989) 1251}
\bibitem{}{S. Albergo {\it{et al}}: {\it{Nucl.Phys.}} {\bf{A590}} (1995) 549c}
\bibitem{}{T. Abbott {\it{et al}}: {\it{Phys.Rev.Lett.}} {\bf{70}} (1993) 1393}
\bibitem{}{J. Barrette {\it{et al}}: {\it{Phys.Rev.Lett.}} {\bf{73}} (1994) 2532}
\bibitem{}{P. Siemens and J. Rasmussen: {\it{Phys.Rev.Lett.}} {\bf{42}} (1979) 880}
\bibitem{}{S. Bass {\it{et al}}: {\it{Phys.Rev.}} C{\bf{51}} (1995) 3343}
\bibitem{}{B.A. Li, W. Bauer and G. Bertsch: {\it{Phys.Rev.}}
C{\bf{44}} (1991) 2095}
\bibitem{}{B.A. Li: {\it{Nucl. Phys.}} A{\bf{570}} (1994) 797}
\end{thebibliography}
\end{document}